\pdfoutput=1

\documentclass[11pt,twoside,a4paper,cmspaper,final,collab]{cms-tdr}
\def\svnVersion{e5eee43-D}\def\svnDate{2019/09/25}\def\cmsCernNoTag{CERN-EP-2019-083}\def\cmsCernDate{\today}\def\cmsMessage{"Published in Physical Review Letters as \href{http://dx.doi.org/10.1103/PhysRevLett.123.131802}{\doi{10.1103/PhysRevLett.123.131802}.}"}

\begin{document}\cmsNoteHeader{HIG-18-020}

\hyphenation{had-ron-i-za-tion}
\hyphenation{cal-or-i-me-ter}
\hyphenation{de-vices}
\RCS$HeadURL$
\RCS$Id$

\newlength\cmsTabSkip\setlength\cmsTabSkip{1ex}
\newlength\cmsFigWidth
\ifthenelse{\boolean{cms@external}}{\setlength\cmsFigWidth{0.48\textwidth}}{\setlength\cmsFigWidth{0.6\textwidth}}
\ifthenelse{\boolean{cms@external}}{\providecommand{\cmsLeft}{upper\xspace}}{\providecommand{\cmsLeft}{left\xspace}}
\ifthenelse{\boolean{cms@external}}{\providecommand{\cmsRight}{lower\xspace}}{\providecommand{\cmsRight}{right\xspace}}
\providecommand{\cmsTable}[1]{\resizebox{\textwidth}{!}{#1}}
\ifthenelse{\boolean{cms@external}}{\providecommand{\CL}{C.L.\xspace}}{\providecommand{\CL}{CL\xspace}}
\ifthenelse{\boolean{cms@external}}{\providecommand{\NA}{\ensuremath{\cdots}\xspace}}{\providecommand{\NA}{\ensuremath{\text{---}}\xspace}}
\ifthenelse{\boolean{cms@external}}{\providecommand{\suppMaterial}{the supplemental material~\cite{SuppMat}}}{\providecommand{\suppMaterial}{Appendix~\ref{app:supp_material}}}

\newcommand{\PP}{\ensuremath{\Pp\Pp}}
\newcommand{\PHp}{\ensuremath{\PH^{+}}}
\newcommand{\emm}{\ensuremath{\Pe\Pgm\Pgm}}
\newcommand{\mmm}{\ensuremath{\Pgm\Pgm\Pgm}}
\newcommand{\Amm}{\ensuremath{\PSA\to\MM}}
\newcommand{\HWA}{\ensuremath{\PH^{+}\to\PWp\PSA}}
\newcommand{\tbH}{\ensuremath{\cPqt\to\cPqb\PH^{+}}}
\newcommand{\BAmm}{\ensuremath{\mathcal{B}(\Amm)}}
\newcommand{\BHWA}{\ensuremath{\mathcal{B}(\HWA)}}
\newcommand{\BtbH}{\ensuremath{\mathcal{B}(\tbH)}}
\newcommand{\XsecTT}{\ensuremath{\sigma(\ttbar)}}
\newcommand{\mHp}{\ensuremath{\mathit{m}_{\PHp}}}
\newcommand{\mA}{\ensuremath{\mathit{m}_{\PSA}}}
\newcommand{\mW}{\ensuremath{\mathit{m}_{\PW}}}
\newcommand{\Mmumu}{\ensuremath{\mathit{m}_{\Pgm\Pgm}}}
\newcommand{\MTW}{\ensuremath{\mT(\ptvec^{\,\Pgm},\ptvecmiss)}}
\newcommand{\ptvecl}{\ensuremath{\ptvec^{\,\Pgm}}}
\newcommand{\absdxy}{\ensuremath{\abs{d_{0}}}}
\newcommand{\absdz}{\ensuremath{\abs{d_{z}}}}
\newcommand{\Irel}{\ensuremath{I_{\text{rel}}}}
\newcommand{\tbHWAMM}{\ensuremath{\tbH\to\cPqb\PWp\PSA\to\cPqb\PWp\MM}}
\newcommand{\Bsig}{\ensuremath{\mathcal{B}_{\text{sig}}}}

\cmsNoteHeader{HIG-18-020}
\title{Search for a light charged Higgs boson decaying to a \texorpdfstring{$\PW$ boson and a CP-odd Higgs boson in final states with $\emm$ or $\mmm$ in proton-proton collisions at $\sqrt{s}=13\TeV$}{W boson and a CP-odd Higgs boson in final states with e mu mu or mu mu mu in proton-proton collisions at sqrt(s) = 13 TeV}}

\date{\today}
\abstract{A search for a light charged Higgs boson ($\PHp$) decaying to a $\PW$ boson and a CP-odd Higgs boson ($\PSA$) in final states with $\emm$ or $\mmm$ is performed using data from $\PP$ collisions at $\sqrt{s}=13\TeV$, recorded by the CMS detector at the LHC and corresponding to an integrated luminosity of 35.9\fbinv. In this search, it is assumed that the $\PHp$ boson is produced in decays of top quarks, and the $\PSA$ boson decays to two oppositely charged muons. The presence of signals for $\PHp$ boson masses between 100 and 160\GeV and $\PSA$ boson masses between 15 and 75\GeV is investigated. No evidence for the production of the $\PHp$ boson is found. Upper limits at 95\% confidence level are obtained on the combined branching fraction for the decay chain, $\tbHWAMM$, of $1.9\times10^{-6}$ to $8.6\times10^{-6}$, depending on the masses of the $\PHp$ and $\PSA$ bosons. These are the first limits for these decay modes of the $\PHp$ and $\PSA$ bosons.}

\hypersetup{
pdfauthor={CMS Collaboration},
pdftitle={Search for a light charged Higgs boson decaying to a W boson and a CP-odd Higgs boson in final states with e mu mu or mu mu mu in proton-proton collisions at sqrt(s) = 13 TeV},
pdfsubject={CMS},
pdfkeywords={CMS, charged Higgs boson, CP-odd Higgs boson, Top}
}

\maketitle

The boson with a mass of 125\GeV discovered in 2012~\cite{Aad:2012tfa,Chatrchyan:2012xdj,Chatrchyan:2013lba} is compatible with the Higgs boson predicted by the standard model (SM)~\cite{Khachatryan:2016vau,Sirunyan:2018koj}. However, this particle can also play the role of a Higgs boson in an extended Higgs sector, which is predicted in many new physics scenarios addressing the hierarchy problem~\cite{Haber:1984rc}, CP violation~\cite{Turok:1990in}, or the mass of neutrinos~\cite{Gelmini:1980re}. As an example, in two Higgs doublet models (2HDMs)~\cite{Lee:1973iz,Branco:2011iw} the 125\GeV boson can be one of the two CP-even Higgs bosons; this class of models also foresees one CP-odd (\PSA) and two charged ($\PH^{\pm}$) Higgs bosons. The observation of additional Higgs bosons would be a clear indication of physics beyond the SM.

We search for an $\PHp$ boson, produced in the decay of a top quark, and decaying to a $\PWp$ boson and an $\PSA$ boson in proton-proton ($\PP$) collisions ($\Pp\Pp\to\ttbar\to\bbbar\PHp\PWm$ and $\HWA$). The charge-conjugated decays are implied throughout this Letter. This production and decay mode of the $\PHp$ boson can be the most dominant one at the LHC if the $\PHp$ boson is lighter than the top quark~\cite{Rathsman:2012dp,Dermisek:2012cn,Kling:2015uba,Arhrib:2016wpw}. This is the first search of this kind at the LHC. The decay mode $\HWA$ in top quark pair events for the mass range $\mW<\mHp<\mathit{m}_{\cPqt}-\mathit{m}_{\cPqb}$ has been studied by the CDF Collaboration assuming that the $\PSA$ boson decays to $\bbbar$ or $\TT$ within specific benchmark scenarios~\cite{Abulencia:2005jd,Aaltonen:2011aj}. The LEP experiments searched for pair production of $\PHp$ bosons in the decay mode $\PHp\to\PW^{+(*)}\PSA$ with $\PSA\to\bbbar$, where an accessible mass range was $\mHp\lesssim100\GeV$~\cite{Abdallah:2003wd,Abbiendi:2008aa,Abbiendi:2013hk}.

In this Letter, we consider ranges of $\mA$ from 15 to 75\GeV and $\mHp$ from ($\mA+85\GeV$) to 160\GeV. The transverse momenta (\PT) of the \PSA boson decay products in this mass region are typically as low as 10--40\GeV. We target the $\Amm$ decay mode, as the use of muons at this energy scale has advantages over using jets or $\Pgt$ leptons in terms of identification efficiency, momentum resolution, and robustness against the number of additional $\PP$ collisions in a single bunch crossing (pileup)~\cite{Sirunyan:2017ulk,Sirunyan:2018fpa,Cacciari:2007fd}. Even though the branching fraction of the \PSA boson, $\BAmm$, is expected to be small (${\lesssim}10^{-3}$) in models such as 2HDMs with a softly broken $\mathbb{Z}_2$ symmetry~\cite{Aoki:2009ha}, the experimental advantages offer a unique opportunity to probe the $\HWA$ decay.

For the {\PW} bosons, the decay modes $\PW\PW\to\ell\Pgn\cPq\cPaq^{\prime}$ ($\ell=\Pe$ or $\Pgm$) are considered. The major background for this search is $\ttbar$ production with at least one lepton originating from jets. Because of the poor resolution of the reconstructed $\mHp$ in the probed mass region, the presence of an excess is investigated in the $\MM$ invariant mass distribution. The search is performed using the $\Pp\Pp$ collision data at $\sqrt{s}=13\TeV$ recorded by the CMS detector at the LHC in 2016. The corresponding integrated luminosity is 35.9\fbinv.

The central feature of the CMS apparatus is a superconducting solenoid of 6\unit{m} internal diameter, providing a magnetic field of 3.8\unit{T}. Within the solenoid volume are a silicon pixel and strip tracker, a lead tungstate crystal electromagnetic calorimeter, and a brass and scintillator hadron calorimeter, each composed of a barrel and two end cap sections. Forward calorimeters extend the pseudorapidity ($\eta$) coverage provided by the barrel and end cap detectors. Muons are detected in gas-ionization chambers embedded in the steel flux-return yoke outside the solenoid. A more detailed description of the CMS detector, as well as definitions of the coordinate system used, can be found in Ref.~\cite{Chatrchyan:2008aa}.

The reconstructed vertex with the largest value of summed $\pt^2$ of physics objects is taken to be the relevant primary $\Pp\Pp$ interaction vertex~\cite{CMSCollaboration:2015zni}. The physics objects are the track-based jets, clustered using the anti-\kt algorithm with a distance parameter of 0.4~\cite{Cacciari:2008gp,Cacciari:2011ma} and the tracks assigned to the vertex as inputs, and the associated track-based missing transverse momentum, taken as the negative vector \pt sum of those jets.

The global event reconstruction is based on the particle-flow algorithm~\cite{Sirunyan:2017ulk}. The algorithm aims to reconstruct and identify each individual particle in an event, with an optimized combination of information from the various elements of the CMS detector. The reconstructed particles are classified as either photons, electrons, muons, or charged or neutral hadrons.

The reconstructed leptons (electrons or muons) are discriminated from nonprompt leptons using tight identification criteria. Nonprompt leptons refer to leptons originating from decays of hadrons or hadrons misidentified as leptons, and prompt leptons refer to leptons from decays of \PW, \PZ, and \PSA bosons, which also include leptons from decays of $\Pgt$ leptons originating from these bosons. Semileptonic decays of $\PB$ hadrons inside jets are the major source of nonprompt leptons in this search. Electron candidates with $\PT>25\GeV$ and within the tracker coverage ($\abs{\eta}<2.5$), excluding the gap between the barrel and end cap calorimeters ($1.44<\abs{\eta}<1.57$), are identified using a multivariate method trained with the track and calorimetric features of electrons as inputs~\cite{Khachatryan:2015hwa}. Electrons from photon conversions are rejected using the information on missing hits in the innermost layers of the tracker and the quality of a fit to a conversion vertex~\cite{Khachatryan:2015hwa}. Muon candidates with $\PT>10\GeV$ within the coverage of the muon detector system ($\abs{\eta}<2.4$) are considered for further identification. For the muon tracks, requirements are placed on the number of hits in the pixel detector, the strip tracker, the muon spectrometer, and the quality of the muon track fit~\cite{Sirunyan:2018fpa}. The transverse (longitudinal) impact parameters $\absdxy$ ($\absdz$) of leptons~\cite{Khachatryan:2015hwa,Sirunyan:2018fpa} are required to be less than 0.25 (1.0)\unit{mm} for electrons and 0.1 (0.5)\unit{mm} for muons, and the $\absdxy$ value divided by its uncertainty is required to be less than 4 for both lepton flavors. The isolation ($\Irel$), which is defined as the ratio of the scalar \PT sum of hadrons and photons, within a cone of ${\Delta}R=\sqrt{\smash[b]{(\Delta\eta)^{2}+(\Delta\phi)^{2}}} < 0.3$ (0.4) around a lepton, to the lepton \PT, is required to be at most 0.06 (0.20) for electrons (muons) after correcting for the contribution from pileup~\cite{Khachatryan:2015hwa,Sirunyan:2018fpa}.

Apart from these tight identification criteria, loose identification criteria are used for the estimation of background from control samples in data and as a part of the jet and lepton veto criteria. For these purposes, relaxed criteria on the \PT of electrons ($>$10\GeV), the multivariate discriminant of electrons, the muon track fit, the impact parameters of muons ($\absdxy<2\unit{mm}$, $\absdz<1\unit{mm}$), and the isolation of electrons (muons) [$\Irel<0.4$ (0.6)] are imposed.

The reconstructed particles are used to form jets and the missing transverse momentum \ptvecmiss. Charged hadrons incompatible with the primary vertex are not considered in the jet reconstruction, and the average neutral pileup contribution is subtracted from the jets~\cite{Sirunyan:2017ulk}. Jets with $\PT>25\GeV$ within $\abs{\eta}<2.4$, which are not in close proximity to any loosely identified lepton [${\Delta}R(\mathrm{j},\ell)>0.4$], are used in this search. The jets originating from {\cPqb} quarks are identified using the combined secondary vertex algorithm v2~\cite{Sirunyan:2017ezt}, and they are referred to as {\cPqb}-tagged jets. The used working point assures an identification efficiency (misidentification probability) of ${\simeq}63\,(1)\%$ for jets originating from {\cPqb} quarks ({\cPqu}, {\cPqd}, {\cPqs} quarks or gluons). The \ptvecmiss is defined as the negative vector \PT sum of all the reconstructed particles in an event~\cite{Sirunyan:2017ulk}.

The search is performed using events with two oppositely charged muons and one additional lepton (electron or muon) in the final state. Signal candidate events are first selected using dilepton triggers~\cite{Khachatryan:2016bia}. Electron-dimuon events are selected by triggers that require the presence of an electron with $\PT>23\GeV$ and a muon with $\PT>8\GeV$. Trimuon events are selected by triggers that require the presence of two muons with $\PT>17$ (8)\GeV for the leading (subleading) muon. The trigger requirements on the leading (subleading) lepton target the lepton from the {\PW} (\PSA) boson. To ensure that the candidate events pass the trigger requirements, an offline condition of $\PT>20\GeV$ is required for the leading muon in trimuon events. Events with exactly three leptons passing the tight identification criteria are used in the search, and events with additional loosely identified leptons are rejected.

Additional conditions are imposed on the candidate events to reduce background contributions. All oppositely charged muon pairs in each event should have an invariant mass satisfying $\Mmumu>12\GeV$ and $\abs{\Mmumu-\mathit{m}_{\PZ}}>10\GeV$ to suppress background processes from the decays of vector mesons or \PZ bosons. At least two jets, of which at least one is \cPqb tagged, are required to remove background contributions not involving {\cPqb} quarks. The remaining background events are expected to be mostly from the \ttbar production where at least one nonprompt lepton originates from a jet, with small contributions from other SM processes involving $\PW/\PZ$ bosons or photon conversions.

The invariant mass of two oppositely charged muons is used to reconstruct the \PSA boson signal. In trimuon events, this muon pair is selected using the muon \PT and the transverse mass, defined as $\MTW=\sqrt{\smash[b]{2(\abs{\ptvecl}\abs{\ptvecmiss}-\ptvecl\cdot\ptvecmiss)}}$, where $\ptvecl$ is the transverse momentum of the muon. The values of \PT and $\MTW$ are typically lower for muons from \PSA bosons than from {\PW} bosons. Among two same-charge muons in the trimuon events, the muon with lower (higher) \PT is assigned to the \PSA ({\PW}) boson. However, if the difference in \PT between these two muons is smaller than 25\GeV and only one of them satisfies $50<\MTW<120\GeV$, consistent with that of a muon from a {\PW} boson, then the other muon is assigned to the \PSA boson. The muons are correctly assigned to their true origins in 59\%--84\% of the events, depending on the $\mA$ and $\mHp$ values. The variation of the efficiency to correctly assign a muon to its mother boson is mainly due to the variation of the \PT of muons from \PSA bosons, and the efficiency is lowest when $\mA\approx\mW$.

Signal processes are modeled at leading order (LO) in quantum chromodynamics (QCD) with \MGvATNLO v2.4.2~\cite{Alwall:2014hca}. As the branching fraction $\BtbH$ is not expected to be large~\cite{Chatrchyan:2013faa,Khachatryan:2015fwh,Khachatryan:2016yzq}, the decay channel $\ttbar\to\bbbar\PH^{+}\PH^{-}$ is not considered in the simulation. All possible decay channels of the two {\PW} bosons are allowed, except processes where both bosons decay to quarks, and the corresponding branching fractions are taken from Ref.~\cite{Tanabashi:2018oca}. The mass of the top quark is set to 172.5\GeV. Since the widths of the $\PHp$ and \PSA bosons are expected to be small in many scenarios~\cite{Aoki:2009ha}, we set their widths to 1\MeV. The width value is less than 1\% of the detector resolution for an $\Mmumu$ value which varies between 0.15 and 0.88\GeV in the range of $\mA$ considered. Signal processes are simulated for 28 mass points in the search region, and the selection efficiencies at intermediate mass values are determined by interpolation.

{\tolerance=800 Backgrounds containing at least three prompt leptons or two prompt leptons and one lepton from a photon conversion are estimated from simulation. Diboson processes ($\PW\cPZ$/$\cPZ\cPZ$), and SM Higgs boson processes with and without $\ttbar$ are simulated at next-to-leading order (NLO) in QCD with \POWHEG v2~\cite{Nason:2004rx,Frixione:2007vw,Alioli:2008tz,Nason:2009ai,Alioli:2010xd,Nason:2013ydw,Hartanto:2015uka}. Other processes are simulated using \MGvATNLO. The $\ttbar\PW$ and $\ttbar\PZ$ processes are simulated at LO precision in QCD with up to two additional partons and the MLM jet merging algorithm~\cite{Alwall:2007fs}. Production of $\cPqt\PZ$, triboson ($\PW\PW\PW$/$\PW\PW\cPZ$/$\PW\cPZ\cPZ$/$\cPZ\cPZ\cPZ$), $\PZ\cPgg$, and $\cPqt\cPaqt\cPgg$ is simulated at NLO precision in QCD with up to one additional parton and the FxFx jet merging algorithm~\cite{Frederix:2012ps}. The background processes are normalized using theoretical cross sections at next-to-next-to-leading order (NNLO) or NLO in QCD~\cite{Cascioli:2014yka,Grazzini:2016swo,Alwall:2014hca,deFlorian:2016spz}. In the case of the $\cPZ\cPgg$ process, the normalization factor is measured using a control sample of data events~\cite{Sirunyan:2017qkz}.\par}

For both the signal and background simulations, the NNPDF3.0 set is used for parton distribution functions (PDFs)~\cite{Ball:2014uwa}. Pileup interactions, parton showers, and hadronization are simulated with \PYTHIA8.212~\cite{Sjostrand:2014zea}, and the simulation of the underlying event is tuned with CUETP8M1~\cite{Khachatryan:2015pea}. All simulated events are passed through the \GEANTfour-based CMS detector simulation~\cite{Agostinelli:2002hh}.

The background yields from processes involving at least one nonprompt lepton from a jet (nonprompt background) are estimated with the tight-to-loose ratio method~\cite{Khachatryan:2016kod}. The method estimates the nonprompt background by applying extrapolation factors on the events with leptons failing the tight identification criteria but passing the loose identification criteria. The extrapolation factors are calculated from the probability of loosely identified leptons from jets to pass the tight identification (tight-to-loose ratio), measured using a control sample enriched with QCD multijet events containing a nonprompt lepton. The tight-to-loose ratio varies between 0.11 and 0.39 (0.071 and 0.14) for muons (electrons), depending on the \PT, $\eta$, and $\Irel$ of the lepton. The validity of the background estimation is verified in samples enriched with nonprompt leptons from $\ttbar$ events and a simulated sample of $\ttbar$ events with at least one nonprompt lepton.

The presence of a signal in the $\Mmumu$ distribution is inspected by comparing the event yield in the data to that of the estimated backgrounds, within a mass window specific to each value of $\mA$. The predicted background distribution in the mass window and sidebands, defined as a mass range between the window edge and up to 5\GeV away from the window center, is approximated by a linear function to estimate the background yield in each signal mass window. The widths ($w$) of the mass windows are optimized to maximize the median significance, $\sqrt{\smash[b]{2[(n_{\mathrm{s}}+n_{\mathrm{b}})\ln(1+n_{\mathrm{s}}/n_{\mathrm{b}})-n_{\mathrm{s}}]}}$~\cite{Cowan:2010js}, where $n_{\mathrm{s}}$ and $n_{\mathrm{b}}$ are the numbers of expected signal and background events in the window. The assumed signal rate for this criterion barely affects the optimization. The width of the signal window is optimized in intervals of 10\GeV. Within each interval, the windows are positioned in steps of 0.45--1.15\GeV with linearly increasing widths. Each window is assigned an index from 1 to 95, increasing with the value of $\Mmumu$ at the window center, as shown in Table~\ref{table:MACuts}.

\begin{table*}[htbp]
  \centering
  \topcaption{Summary of mass windows ($\abs{\Mmumu-\mA}<w$) for each $\mA$ hypothesis.}
  \cmsTable{
    \small\begin{scotch}{ l  c  c  c  c  c  c  c}
    $\mA$ range ({\GeVns}) & $[15,25)$   & $[25,35)$   & $[35,45)$   & $[45,55)$   & $[55,65)$   & $[65,75)$   & 75  \\
    Window index           & 1--23       & 24--42      & 43--59      & 60--73      & 74--85      & 86--94      & 95  \\
    $\mA$ step ({\GeVns})  & 0.45        & 0.55        & 0.6         & 0.75        & 0.9         & 1.15        & \NA \\
    $w$ ({\GeVns})         & $[0.5,0.7)$ & $[0.7,0.8)$ & $[0.8,1.0)$ & $[1.0,1.2)$ & $[1.2,1.5)$ & $[1.5,1.8)$ & 1.8 \\
    $\Mmumu$ resolution ({\GeVns}) & $[0.15,0.28)$ & $[0.28,0.40)$ & $[0.40,0.52)$ & $[0.52,0.64)$ & $[0.64,0.76)$ & $[0.76,0.88)$ & 0.88 \\
    \end{scotch}
  }
\label{table:MACuts}
\end{table*}

\begin{figure}[htbp]
\centering
  \includegraphics[width=\cmsFigWidth]{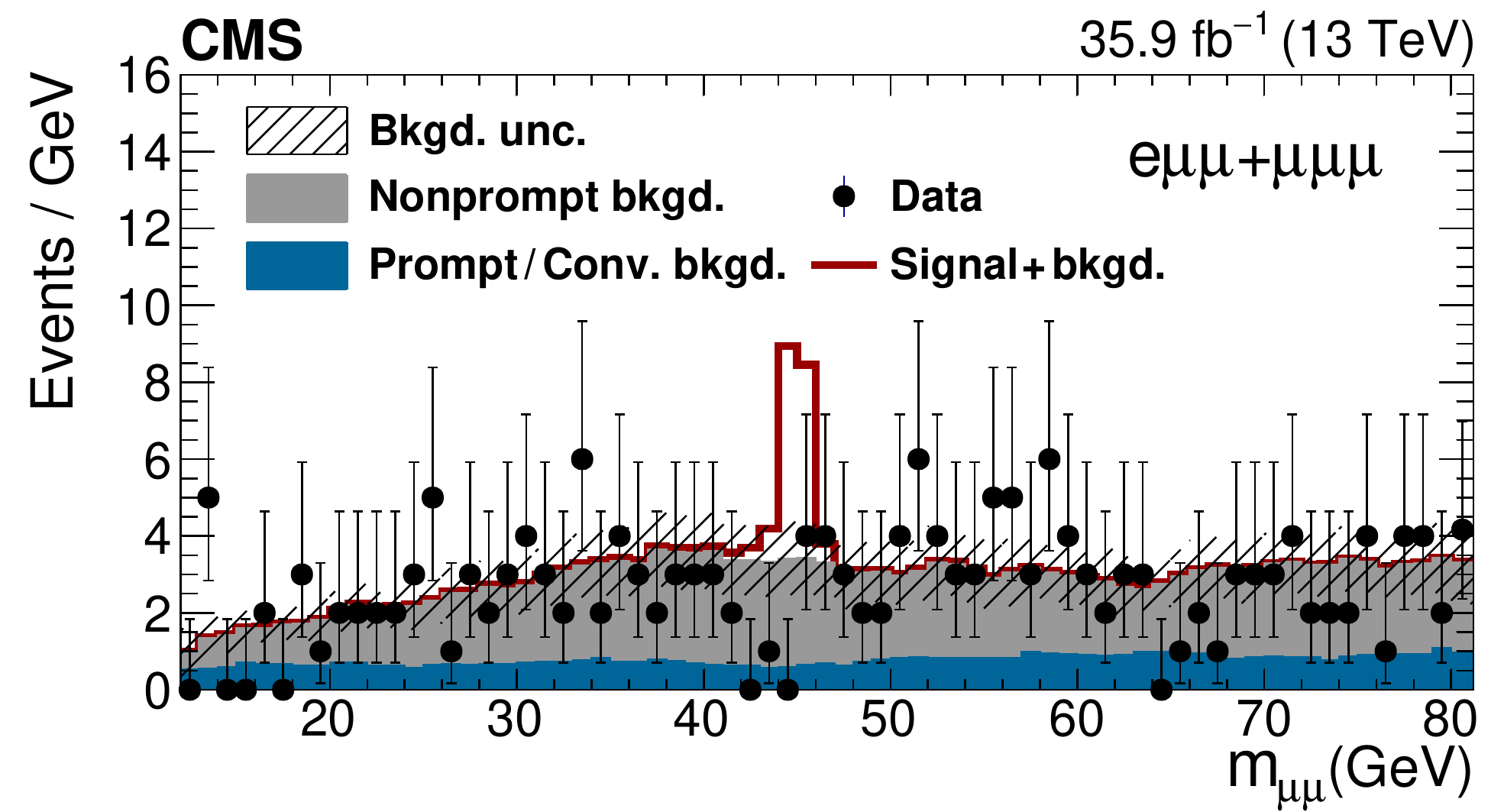}
  \caption{The $\Mmumu$ distribution of candidate muon pairs from \PSA bosons in the $\emm$ and $\mmm$ final states. A constant bin size (1\GeV) is used in the figure except for the last bin of $[80,\,81.2]$ ({\GeVns}). The expected signal distribution for $\mHp=130$ and $\mA=45\GeV$ is also shown on top of the expected backgrounds assuming $\XsecTT=832\unit{pb}$ and $\BtbH\BHWA\BAmm=6\times10^{-6}$.}
  \label{fig:ResultYield}
\end{figure}

The $\Mmumu$ distribution of candidate muon pairs from \PSA bosons is shown in Fig.~\ref{fig:ResultYield}. The corresponding figures, event yields in the mass windows, and signal efficiencies for individual final states are available in \suppMaterial. In the presence of a signal, an excess in the yield is expected in a narrow range around $\mA$ above a smooth background, as shown for $\mA=45\GeV$ in the figure. Differential event rates of the nonresonant distribution from incorrectly selected pairs are negligible when compared to the resonant part from the correct assignment, leading to the $\Mmumu$ distribution of a signal mainly determined by the $\mA$ value.

No evidence of a signal is observed in the $\Mmumu$ spectrum. Upper limits at 95\% confidence level (\CL) on the product of branching fractions, $\Bsig=\BtbH\BHWA\BAmm$, are set, using the $\CLs$ criterion~\cite{Junk:1999kv,Read:2002hq,ATLAS:2011tau}, based on the combined likelihood of event yields in the mass windows from the $\emm$ and $\mmm$ channels. In the calculation, the \ttbar production cross section is set to the SM prediction of 832\unit{pb}, computed at NNLO in QCD, including soft-gluon resummation to next-to-next-to-leading logarithmic order~\cite{Czakon:2011xx}. The systematic uncertainties are treated as nuisance parameters with a log-normal distribution for their likelihood. Typical magnitudes of an overall uncertainty are 30\% for the backgrounds and 7\% for the signal. The impact of the systematic uncertainties on the result is small because of the large statistical uncertainty of the data.

The largest source of uncertainty arises from the estimation of the nonprompt lepton background, which is determined from both simulation and data. In the simulation, a comparison is performed between the yield of simulated $\ttbar$ events passing the event selection and the calculated yield from the tight-to-loose ratio method applied to the simulated $\ttbar$ sample. The tight-to-loose ratio from simulated multijet events is used in the calculation. The two values agree within 27\% (23\%) in the $\emm$ ($\mmm$) channel. In the data, the dependence of the tight-to-loose ratio arising from uncertainties in the jet energy scale, the flavor of the parton that generates the nonprompt lepton, and the estimation of the prompt lepton contribution in the control sample for the measurement of the tight-to-loose ratio are considered. The first two sources are examined by varying the \PT selection applied to the jets or by requiring the presence of a {\cPqb}-tagged jet in the sample, and the last source is examined by varying the normalization of the residual prompt lepton contribution by its own uncertainty. The impact of each variation on the prediction is observed to be 7\%, 13\%, and 3\%, respectively, and it results in a total variation of 15\%, when added in quadrature. Reflecting the observed differences, a systematic uncertainty of 30\% is assigned for this background.

Subleading sources of uncertainty arise from the limited sample size for the estimation of the nonprompt lepton background (20\%) and from the interpolation used in the determination of signal efficiency (5\%). The systematic uncertainties associated with the modeling of the backgrounds using the sidebands and other experimental and theoretical sources are also examined. However, their magnitude is observed to be negligible compared to those of the aforementioned sources. These include the lepton identification efficiency, trigger efficiency, {\cPqb} tagging efficiency~\cite{Sirunyan:2017ezt}, the energy scale and resolution of leptons and jets~\cite{Khachatryan:2015hwa,Khachatryan:2016kdb,Sirunyan:2018fpa}, the momentum scale of unclustered objects that affects \ptvecmiss~\cite{CMS:2016ljj}, the integrated luminosity measurement~\cite{CMS:2017sdi}, the total inelastic $\Pp\Pp$ cross section that affects the pileup modeling in simulation, the measured normalization factor of $\cPZ\cPgg$ processes, the choice of PDFs, and factorization and renormalization scales that affect the normalization of simulated samples and signal acceptances~\cite{Cascioli:2014yka,Grazzini:2016swo,deFlorian:2016spz,Butterworth:2015oua}.

\begin{figure}[htbp]
\centering
  \includegraphics[width=\cmsFigWidth]{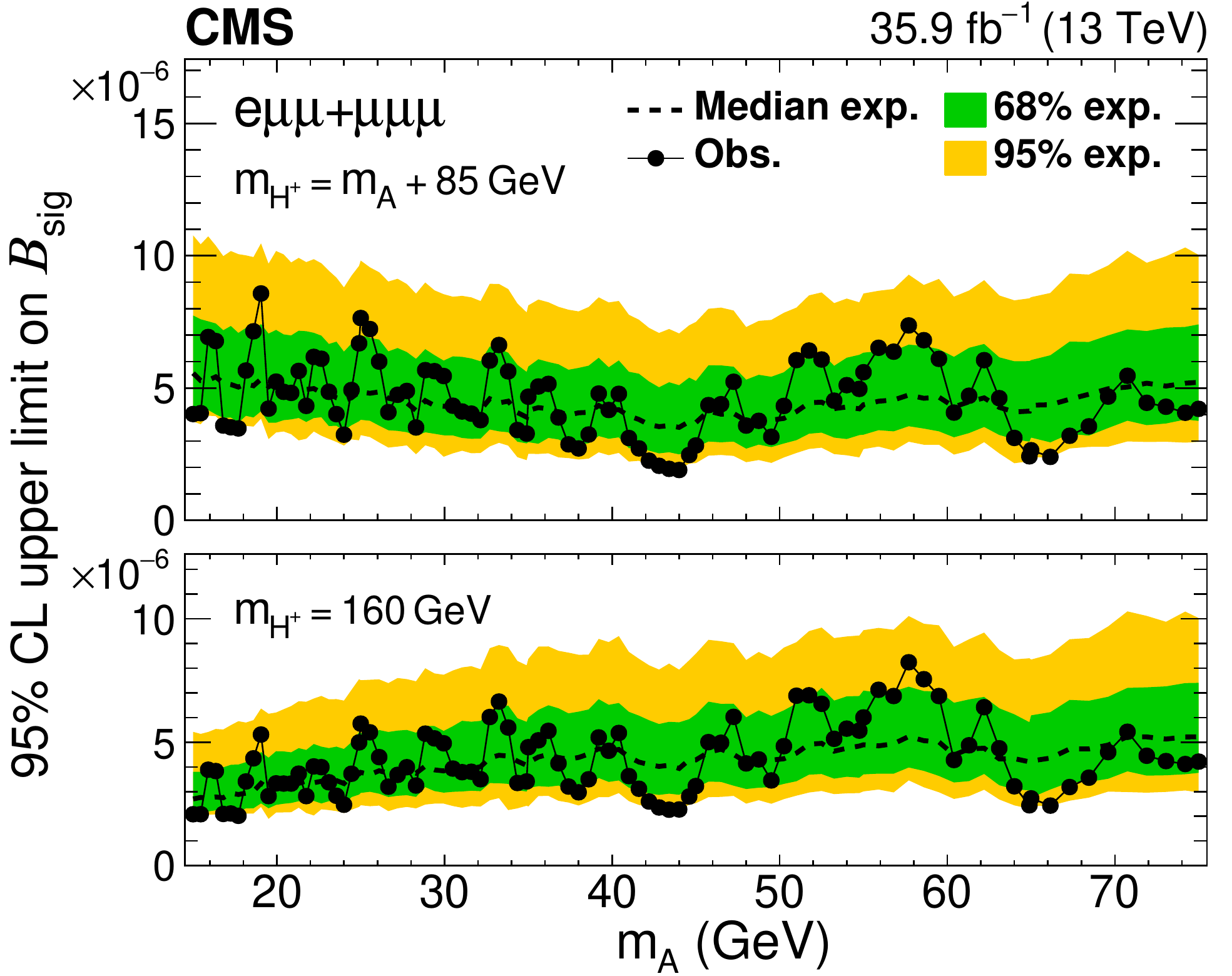}
  \caption{Expected and observed upper limits at 95\% \CL on $\Bsig$ for the $\mA$ values defined in Table~\ref{table:MACuts}, with an assumption of $\mHp=\mA+85\GeV$ (upper) or $\mHp=160\GeV$ (lower). The green (yellow) bands indicate the regions containing 68 (95)\% of the limit values expected under the background-only hypothesis.}
  \label{fig:ResultLimit}
\end{figure}

The expected and observed upper limits on $\Bsig$ for the 95 $\mA$ values defined in Table~\ref{table:MACuts} are shown in Fig.~\ref{fig:ResultLimit}. The corresponding limits for individual final states are available in \suppMaterial. The limits are presented as a function of $\mA$ for two $\PHp$ boson masses, $\mHp=\mA+85\GeV$ and $\mHp=160\GeV$. The difference of the limits for the two $\mHp$ values is smaller than their uncertainties. Short-range bin-to-bin correlations originate from the overlap between neighboring search windows. The observed upper limit on $\Bsig$ varies between $1.9\times10^{-6}$ and $8.6\times10^{-6}$ depending on the assumed values of $\mHp$ and $\mA$, and $\Bsig>8.6\times10^{-6}$ is excluded at $95\%$ \CL in the entire search region. These are the first limits on the combined branching fraction for the decay chain, $\tbHWAMM$. In type-I/II 2HDMs~\cite{Kling:2015uba,Arhrib:2016wpw} or the next-to-minimal supersymmetric SM~\cite{Rathsman:2012dp,Dermisek:2012cn} where $\BAmm\approx3\times10^{-4}$ holds~\cite{Aoki:2009ha,Almarashi:2018ign}, these upper limits on $\Bsig$ impose a constraint $\BtbH\BHWA\lesssim2.9\%$ at 95\% \CL, more stringent than the previous results reported by the CDF Collaboration, using different decay modes of the \PSA boson~\cite{Abulencia:2005jd,Aaltonen:2011aj}. 

In summary, a search is performed for a charged Higgs boson $\PHp$, produced in the decay of a top quark, and decaying further into a $\PW$ boson and a CP-odd Higgs boson $\PSA$, where the $\PSA$ boson decays to two muons. The analysis uses proton-proton collision data at $\sqrt{s}=13\TeV$, recorded by the CMS experiment, corresponding to an integrated luminosity of 35.9\fbinv. A resonant signature in the dimuon mass spectrum is searched in trilepton events for the ranges of $\mA$ between 15 and 75\GeV and $\mHp$ between ($\mA+85\GeV$) and $160\GeV$. No statistically significant excess is found. Upper limits at 95\% confidence level on the product of branching fractions, $\BtbH\BHWA\BAmm$, of $1.9\times10^{-6}$ to $8.6\times10^{-6}$ are obtained, depending on the masses of the $\PHp$ and $\PSA$ bosons. The reported analysis constitutes the first search for the $\HWA$ process in the $\Amm$ decay channel.

\begin{acknowledgments}
We congratulate our colleagues in the CERN accelerator departments for the excellent performance of the LHC and thank the technical and administrative staffs at CERN and at other CMS institutes for their contributions to the success of the CMS effort. In addition, we gratefully acknowledge the computing centers and personnel of the Worldwide LHC Computing Grid for so effectively delivering the computing infrastructure essential to our analyses. Finally, we acknowledge the enduring support for the construction and operation of the LHC and the CMS detector provided by the following funding agencies: BMBWF and FWF (Austria); FNRS and FWO (Belgium); CNPq, CAPES, FAPERJ, FAPERGS, and FAPESP (Brazil); MES (Bulgaria); CERN; CAS, MoST, and NSFC (China); COLCIENCIAS (Colombia); MSES and CSF (Croatia); RPF (Cyprus); SENESCYT (Ecuador); MoER, ERC IUT, PUT and ERDF (Estonia); Academy of Finland, MEC, and HIP (Finland); CEA and CNRS/IN2P3 (France); BMBF, DFG, and HGF (Germany); GSRT (Greece); NKFIA (Hungary); DAE and DST (India); IPM (Iran); SFI (Ireland); INFN (Italy); MSIP and NRF (Republic of Korea); MES (Latvia); LAS (Lithuania); MOE and UM (Malaysia); BUAP, CINVESTAV, CONACYT, LNS, SEP, and UASLP-FAI (Mexico); MOS (Montenegro); MBIE (New Zealand); PAEC (Pakistan); MSHE and NSC (Poland); FCT (Portugal); JINR (Dubna); MON, RosAtom, RAS, RFBR, and NRC KI (Russia); MESTD (Serbia); SEIDI, CPAN, PCTI, and FEDER (Spain); MOSTR (Sri Lanka); Swiss Funding Agencies (Switzerland); MST (Taipei); ThEPCenter, IPST, STAR, and NSTDA (Thailand); TUBITAK and TAEK (Turkey); NASU and SFFR (Ukraine); STFC (United Kingdom); DOE and NSF (USA).
\end{acknowledgments}

\bibliography{auto_generated}

\ifthenelse{\boolean{cms@external}}{}{
\numberwithin{figure}{section}
\clearpage
\appendix
\section{Signal efficiencies, \texorpdfstring{\Mmumu}{M[mumu]} distributions, event yields, and upper limits on \texorpdfstring{\Bsig}{B[sig]} for individual final states\label{app:supp_material}}
\begin{figure*}[htbp]
\centering
  \includegraphics[width=0.48\textwidth]{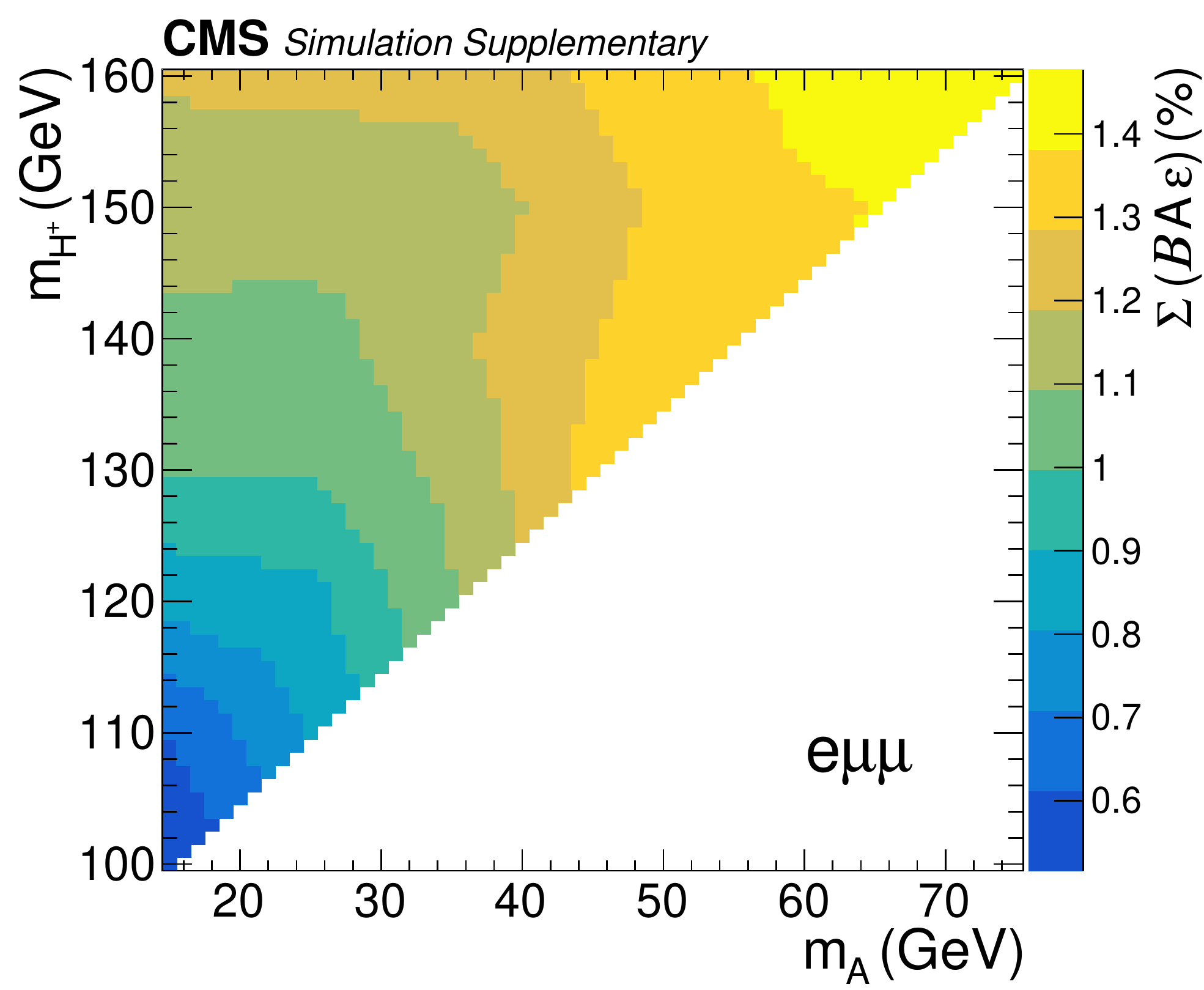}
  \includegraphics[width=0.48\textwidth]{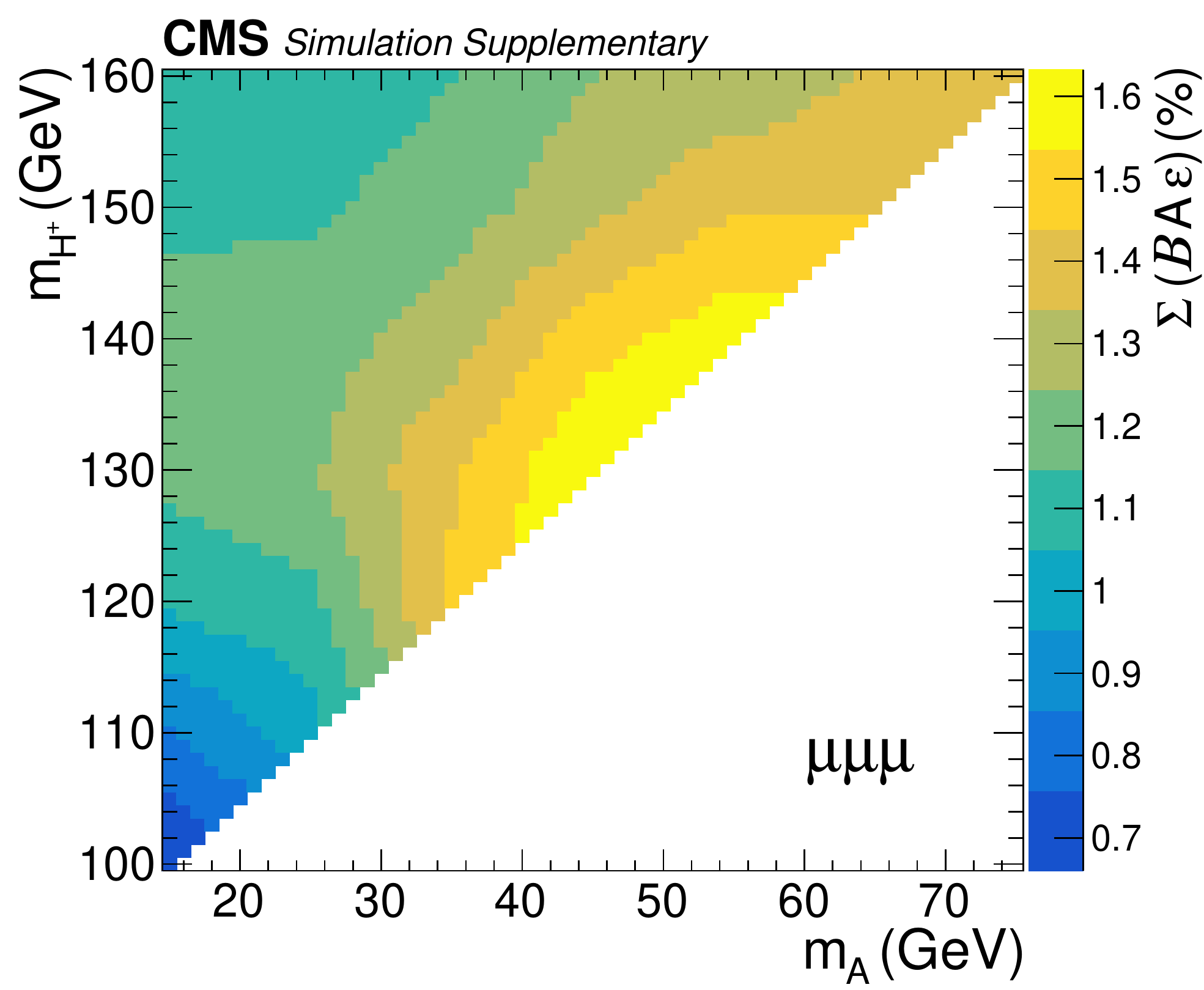}
  \caption{The fraction of signal events passing the final event selection in the $\emm$ (left) and $\mmm$ (right) final states. The fraction is relative to the yield before the decays of the two ${\PW}$ bosons in the signal processes ($\ttbar\to\bbbar\PWp\PWm\MM$), which include the branching fraction of each decay mode of the two ${\PW}$ bosons ($\mathcal{B}$) and the acceptance (A) times efficiency ($\varepsilon$) of the event selection for the decay mode. All decay modes of the two ${\PW}$ bosons are considered in the calculation except the cases where both of the bosons decay hadronically.}
  \label{fig:SigEff}
\end{figure*}

\begin{figure*}[htbp]
\centering
  \includegraphics[width=0.48\textwidth]{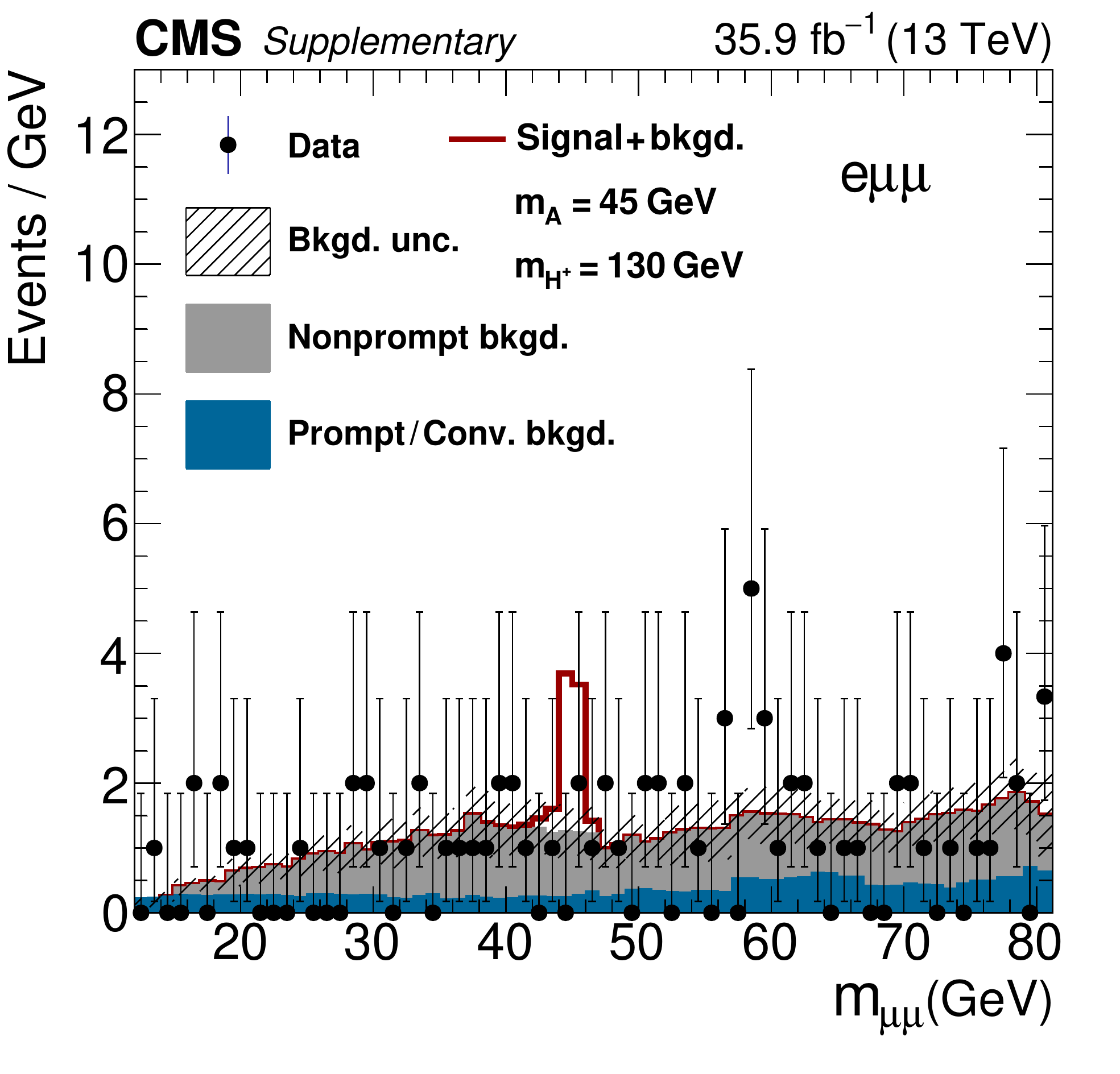}
  \includegraphics[width=0.48\textwidth]{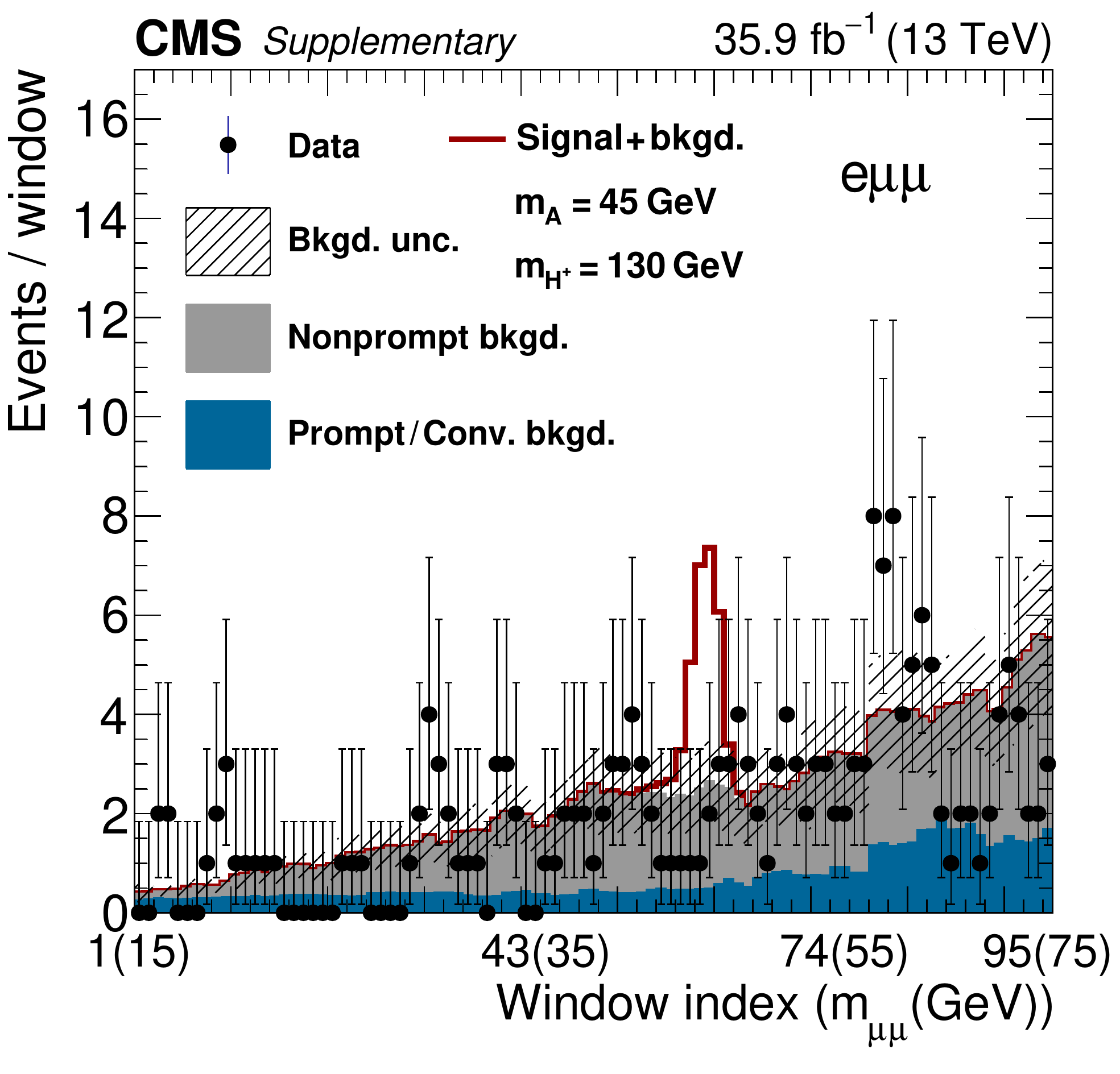}
  \includegraphics[width=0.48\textwidth]{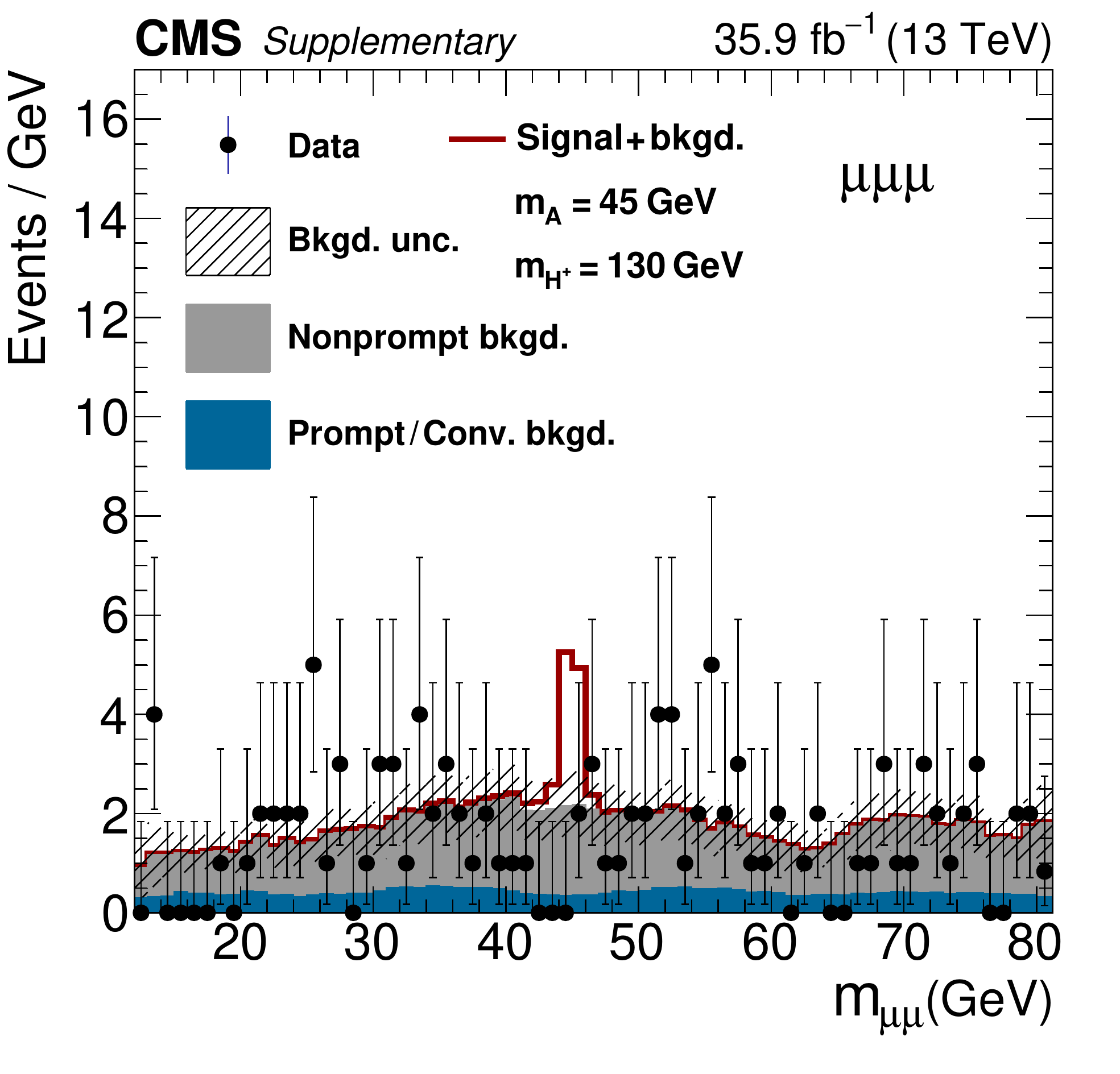}
  \includegraphics[width=0.48\textwidth]{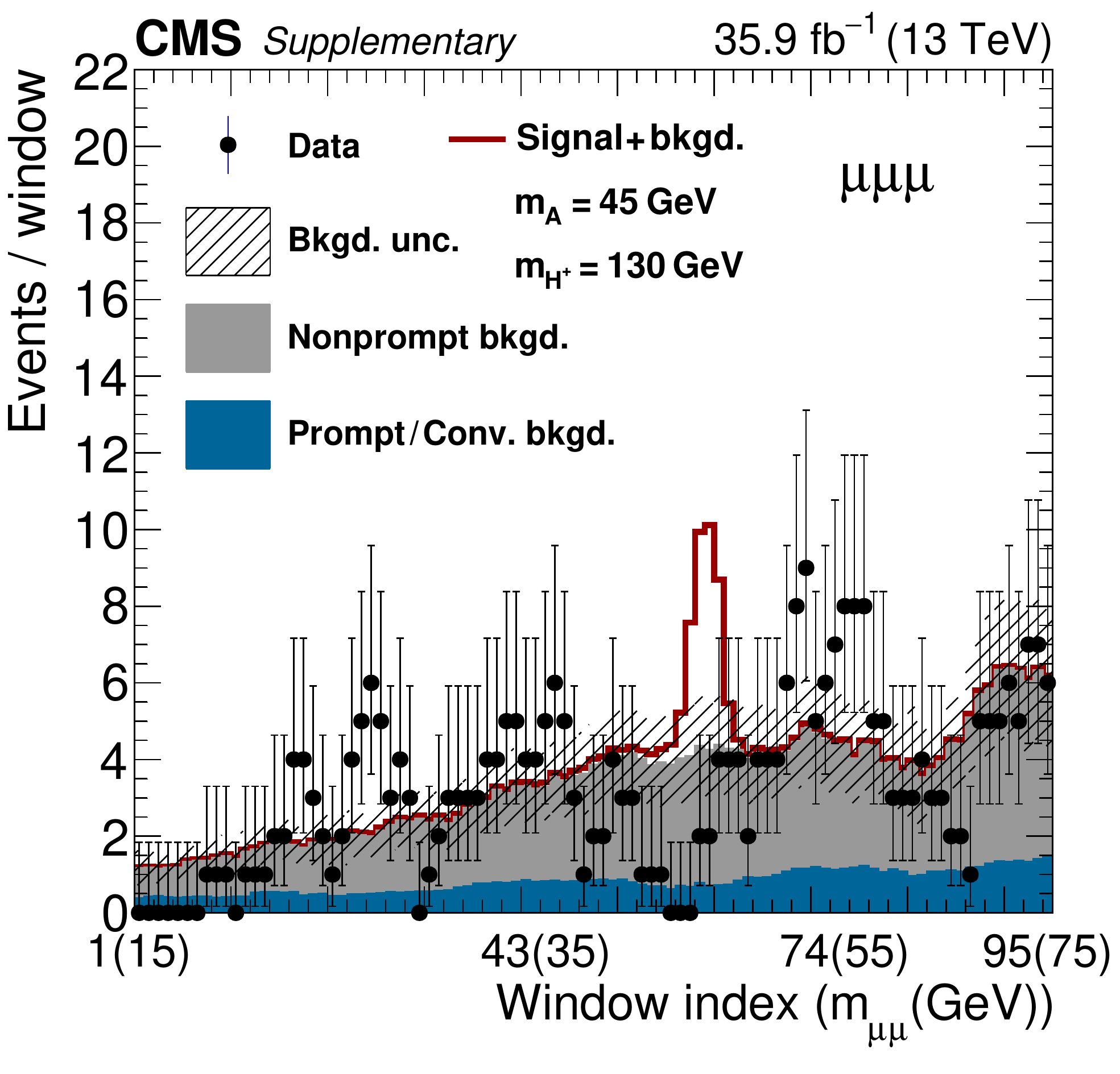}
  \caption{The $\Mmumu$ distribution of candidate muon pairs from \PSA bosons (left) and the event yields in each signal window (right) in the $\emm$ (upper) and $\mmm$ (lower) final states. A constant bin size (1\GeV) is used in the left figures except the last bin of $[80,\,81.2]$ ({\GeVns}). Values of $\Mmumu$ at centers of the corresponding windows are written in the parentheses on the $x$ axis of the right figures. The expected signal distribution for $\mHp=130$ and $\mA=45\GeV$ is also shown on top of the expected backgrounds assuming $\XsecTT=832\unit{pb}$ and $\Bsig=6\times10^{-6}$.}
  \label{fig:MassYieldPlots}
\end{figure*}

\begin{figure*}[htbp]
\centering
  \includegraphics[width=0.48\textwidth]{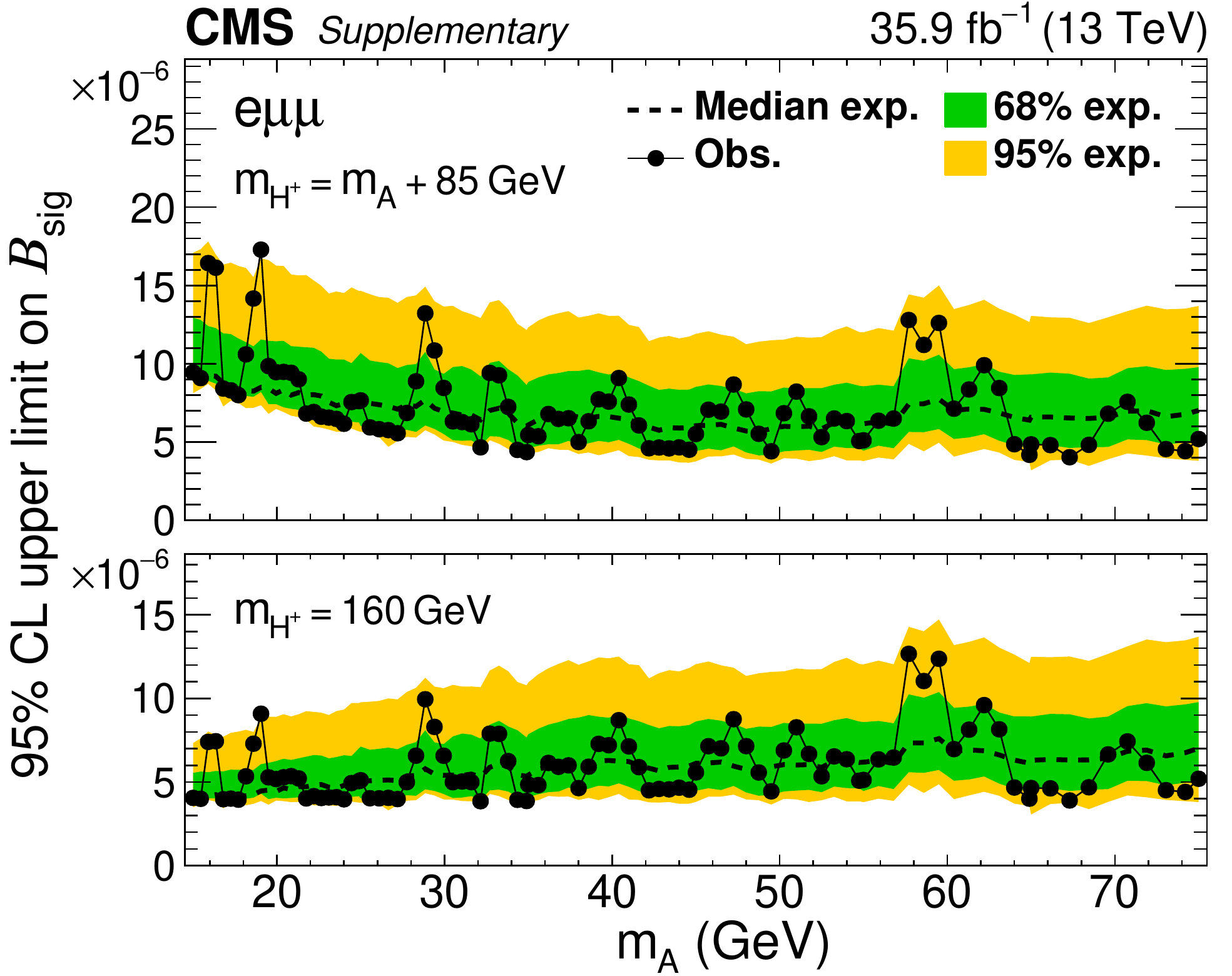}
  \includegraphics[width=0.48\textwidth]{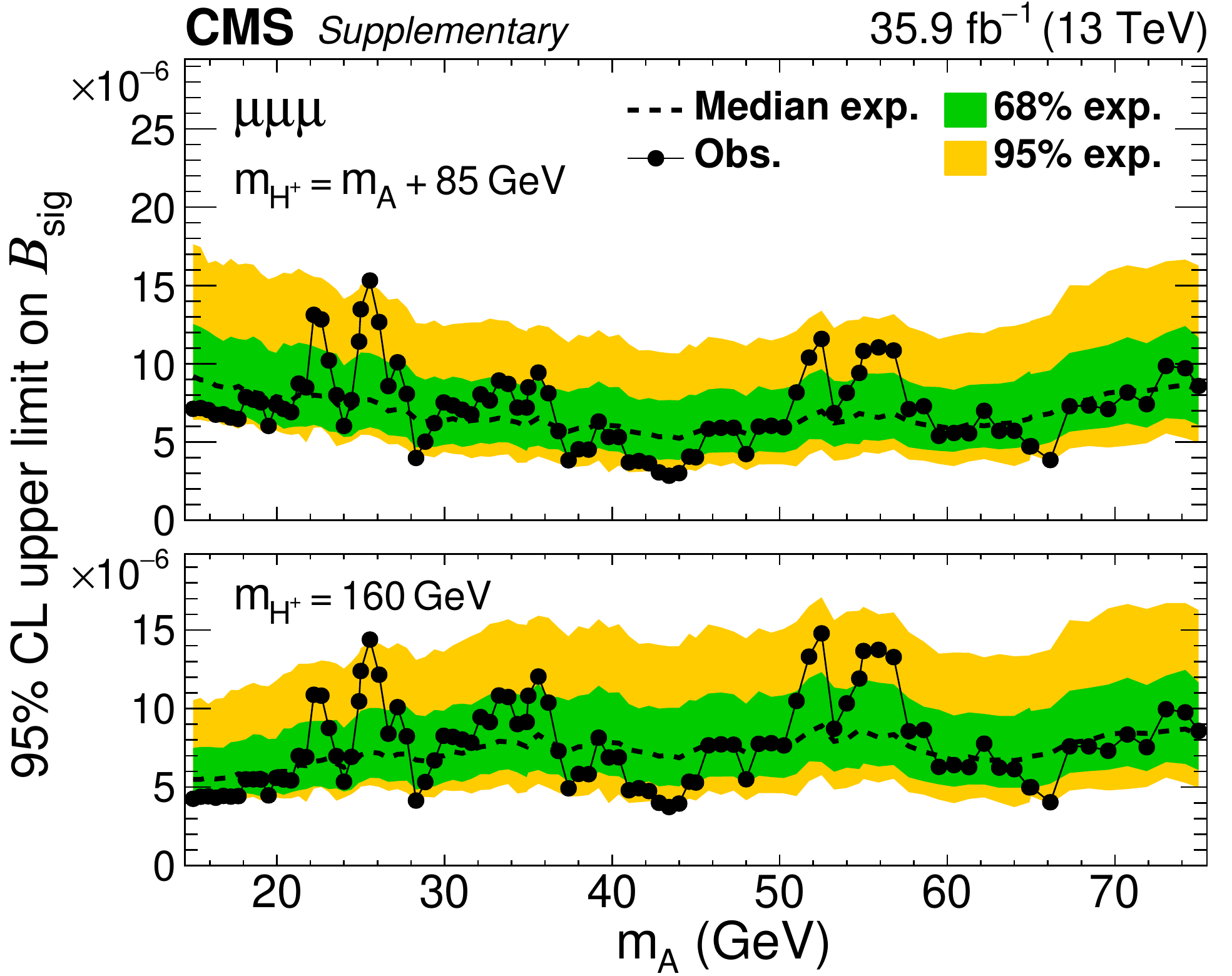}
  \caption{Upper limits at 95\% \CL on $\Bsig$ for the 95 $\mA$ values, with an assumption of $\mHp=\mA+85\GeV$ (upper) or $\mHp=160\GeV$ (lower), for individual final states (left: $\emm$ and right: $\mmm$ final states). In the calculation, the same value of $\XsecTT$ as in Fig.~\ref{fig:MassYieldPlots} is assumed.}
  \label{fig:Lim1DIndividual}
\end{figure*}

}
\cleardoublepage \section{The CMS Collaboration \label{app:collab}}\begin{sloppypar}\hyphenpenalty=5000\widowpenalty=500\clubpenalty=5000\vskip\cmsinstskip
\textbf{Yerevan Physics Institute, Yerevan, Armenia}\\*[0pt]
A.M.~Sirunyan$^{\textrm{\dag}}$, A.~Tumasyan
\vskip\cmsinstskip
\textbf{Institut für Hochenergiephysik, Wien, Austria}\\*[0pt]
W.~Adam, F.~Ambrogi, T.~Bergauer, J.~Brandstetter, M.~Dragicevic, J.~Erö, A.~Escalante~Del~Valle, M.~Flechl, R.~Frühwirth\cmsAuthorMark{1}, M.~Jeitler\cmsAuthorMark{1}, N.~Krammer, I.~Krätschmer, D.~Liko, T.~Madlener, I.~Mikulec, N.~Rad, J.~Schieck\cmsAuthorMark{1}, R.~Schöfbeck, M.~Spanring, D.~Spitzbart, W.~Waltenberger, J.~Wittmann, C.-E.~Wulz\cmsAuthorMark{1}, M.~Zarucki
% ...
\vskip\cmsinstskip
\dag: Deceased\\
1:  Also at Vienna University of Technology, Vienna, Austria\\
% ...
\end{sloppypar}
\end{document}